\documentclass[11pt]{article}
\usepackage{mathptm}
\usepackage{amsmath}
\usepackage{mdwlist}

\newcommand{\locsec}[1]{\bigskip\pagebreak[3]\textbf{#1}\nopagebreak[4]}

\begin{document}

\begin{center}

\textbf{\large Significance Tests in Climate Science}\\[.5em]
Maarten H. P. Ambaum\\
\textsl{Department of Meteorology, University of Reading, UK}\\
(15 March 2010)\\[1em]

\parbox{0.95\textwidth}{A large fraction of papers in the climate literature includes erroneous uses of significance tests.  A Bayesian analysis is presented to highlight the meaning of significance tests and why typical misuse occurs.  It is concluded that a significance test very rarely provides useful quantitative information.  The significance statistic is not a quantitative measure of how confident we can be of the `reality' of a given result.
}
\end{center}

\locsec{1. Introduction}

In the climate literature one can regularly read statements such as `this correlation is 95\% significant' or `areas of significant anomalies at the 90\% significance level are shaded' or `the significant values are printed in bold.' Unfortunately this is an incorrect and misleading way of using significance tests.  In this note we highlight why this is wrong.  We will also indicate what a significance test does mean.

Although this note does not add new theory to significance tests, it does employ a Bayesian framework to exemplify the issues.  Practitioners in climate science are generally familiar with the technical aspects of Bayesian statistics, but will perhaps be less familiar with its use in the analysis of significance tests.  

We tested a recent, randomly selected issue of \textsl{The Journal of Climate} for at least one such misuse of significance tests in each article. \textsl{The Journal of Climate} was not selected because it is prone to include such errors but because it can safely be considered to be one of the top journals in climate science.  In that particular issue we observed a misuse of significance tests in 14 out of 19 articles.  A randomly selected issue of ten years before showed such misuse of significance tests in 7 out of 13 articles.  These two samples perhaps would not pass a traditional significance test, but they do indicate that such errors occur in the best journals with the most careful writing and editing.  Indeed, in one of this author's papers such erroneous use occurred.

Comparing the papers in the two examined issues, it appears that papers with a more dynamical focus generally do not stray as much into significance testing as paper with a more geographical, diagnostic focus.  The distinction between these two categories is necessarily vague.  The author also wonders whether an increased ease with which such tests can be performed with data processing and plotting software has lead to a near default inclusion of such tests in papers.  From experience, the author is also aware that reviewers often insist on the inclusion of significance tests.
%
%

This reported misuse of significance tests does not necessarily invalidate the results from those parts of the papers. The significance test is sometimes only a small part of the evidence presented, often it is only a subsidiary, if misleading piece of information.  Furthermore, many papers contain somewhat neutral statements such as `this correlation is highly significant ($p<0.01$).'  Such a statement could be read at face value, namely that the correlation was subjected to a significance test and a $p$-value of less than 0.01 was found.  In such a neutral reading, the statement is also somewhat meaningless, as will be shown below.

Such a statement is more likely intended to mean that the correlation is in some sense `real' and the $p$-value is a quantification of that.  We will show here that this quantification of confidence is wrong.  Data highlighted as significant may easily be less significant than data that were suppressed as not passing the significance threshold.  Simply put, the significance statistic is not a quantitative measure of how confident we can be of the `reality' of a given result.

A typical scenario in which people use significance tests in climate science is the following: some experiment produces two time-series and they are correlated (for example, global mean temperature and the ENSO index.)  Is the observed correlation real or is it a fluke?  We will use this correlation scenario throughout to be able to exemplify specific aspects of significance testing; however, the discussion is valid for any significance test which is based on assessing the probability of an alternative hypothesis, the null-hypothesis, which assumes that the data exhibit no relation.

So let us concentrate on the typical question of whether an observed correlation is real or a statistical fluke.  The correct answer to this question is in fact very difficult to obtain.  Indeed, it is usually impossible to quantify our degree of believe either way by statistics alone.  Unfortunately, it is widely held that a standard significance test (for example, a $t$-test) provides an answer.  Standard significance tests hardly ever give a useful answer to the question we are trying to answer.

It can be argued that the significance test more accurately should be named the insignificance test, as it may be a reasonable test for insignificance.  Clearly if Fisher had called his test the insignificance test, then it would probably not be used very much.  Marketing plays an important role in science.

There is quite a bit of literature on the misuse of statistical significance tests.  It has been argued that the power of R.~A.~Fisher, the great proponent of significance testing, is the real reason why significance tests are so ubiquitous; see Ziliak and McCloskey (2008).   In the psychological literature the false use of significance tests has been regularly pointed out, although, perhaps, not with much success.  See for example Cohen (1994), Hunter (1997), or Armstrong (2007).  In the geophysical literature there has been much less attention to the misuse of significance tests.  A nice review of significance testing in atmospheric science, including stern critique of the misuse of significance testing, can be found in Nicholls (2001).   A thorough and detailed discussion in the context of scientific hypothesis testing can be found in Jaynes, 2003.

In the next section we will highlight the general structure of a significance test and exemplify, using frequency tables, the relationship between what the significance test provides and what we really would like to know.  Section 3 provides a Bayesian analysis of significance tests.  This quantifies the relationship between significance tests and hypothesis tests.  It also quantifies what we do get out of a significance test.  Some concluding remarks regarding the practical use of significance tests are in section 4.

\locsec{2. General structure of significance tests}

First, let us examine the structure of a typical significance test in the scenario described before.  A brief introduction can also be found in Jolliffe, 2004.  The hypothesis we are trying to test is: `the two time-series are related; the correlation $r_0$ we observe is a measure of this relation.'  Note the distinction between relation and correlation here.  A correlation is a statistical property of two time-series, while a relation indicates that the two time-series are dependent in some physical way.    We then define the so-called \textsl{null-hypothesis} which in some sense states the opposite.  In our case: `The two time-series are not related; the observed correlation $r_0$ is a fluke.'  We then continue to test the validity of the null hypothesis.

Here is where the first confusion comes in.  We want to devise a way to assign a probability to the validity of the null-hypothesis, given the observed correlation.  But what we end up doing is calculating the probability of the observed correlation when the null hypothesis is assumed to be true.  These two probabilities are different, although they are related by Bayes' theorem.  This common error is called the error of the \textsl{transposed conditional.}  The discussion of Bayesian statistics, below, formalizes this.

Let us continue with the usual significance test.  There are standard procedures for assigning a probability to the observed correlation, assuming the null-hypothesis is true: $t$-tests for Gaussian data, parametric or or non-parametric tests (for example, the Kolmogorov--Smirnov test) for non-Gaussian data.  In general, we study synthetic time-series with similar properties, perhaps similar temporal autocorrelation, etc., to the original time-series but which are unrelated by construction.  We can then see what the probability is to find a correlation between such unrelated series at least as large as the observed correlation $r_0$. This probability is called the $p$-value.

There is a distinction between the use of the absolute value of the correlation or the actual value which then correspond to a two-sided or a one-sided test, respectively.  The presented arguments work the same for either test and also wider classes of tests: significance tests always find the probability, the $p$-value, of an observation, assuming the truth of the null-hypothesis.

If the $p$-value is large then two unrelated time-series can easily produce a correlation as large as $r_0$.  We must then conclude that the observed correlation provides little evidence for an actual relation between the two original time-series.  If the $p$-value is low (typically, values of 5\% or even 1\% are chosen to define what is `low') then the observed correlation is unlikely to occur in unrelated time-series.

What can we conclude from those two possible outcomes?  It is reasonable to conclude that, if we only have these statistics available, a high $p$-value is a good indicator that the observed correlation $r_0$ is not particularly special.  Any pair of unrelated time-series could easily (high $p$) have a correlation as large as $r_0$.   Note that this does not mean that the null-hypothesis is highly probable; it means that the correlation value is highly probable, given that the null-hypothesis is true.  Beware of the error of the transposed conditional.  

Further confusion occurs when the $p$-value is low.  All it means is that it not likely that the observed correlation would occur in two unrelated time-series. However, we cannot conclude from such an outcome that the two original time-series are likely related, that is, significantly correlated.  

It can be argued that `significantly correlated' is \textsl{defined} to correspond to a low $p$-value.  Although this would be technically correct, it would render the statement of significant correlation quite insignificant in any practical sense.  The low $p$-value is a property of unrelated time-series; it says nothing about related time-series.  In philosophy such a situation is called a \textsl{category error}.  Statements such as `the two time-series are significantly correlated at the 95\% level' (that is $p$ is lower than 5\%) commit a category error.

It is instructive to work this out using a $2\times2$ frequency table.  Suppose we can repeat our experiment that produced the two original time-series as often as we like and we know beforehand that the series are related.  For example, we run an ensemble of climate models and extract the global mean temperature and the ENSO-signal for each ensemble member.  We then find some correlation $r$ between the two time-series.  We can then compare that correlation with the threshold correlation, say $r_p$, which corresponds to a given $p$-value.   For example we can chose a $p$-value for significance of 5\%.  This will correspond to a particular threshold correlation $r_p.$  The correlation between the related time-series of any experiment will be either larger or smaller than $r_p$.

We have not dwelled on what is meant when we know something to be true beforehand.  In science, we need to use an operational definition stating that there is a wide body of historical evidence which supports the hypothesis.  For example, Netwon's laws are known to be `true.'  This example is so well-known that we immediately can understand the subtleties of scientific truths.  We know for example that Newton's laws have a limited validity.  Scientific truth always has to be qualified; it cannot be compared with logical truth.  A wide-ranging discussion can be found in Jaynes, 2003.

In our example we run a hundred experiments and divide them in two categories with either high (higher than $r_p$) or low (lower than $r_p$) correlation.  Because the time-series are related by construction we expect a fairly large fraction to produce a high correlation.  Let us, for the sake of argument, say that 60\% of our experiments show a high correlation.

We now do the same thing for a hundred synthetic time-series which are unrelated by construction.  If our significance test is defined properly, then out of a hundred unrelated synthetic time-series, on average, 5 will have a high correlation and 95 will have a low correlation.  The results are summarized in the table, below.
\begin{table}[h!]
\begin{center}
\begin{tabular}{r|c|c}
& low $r$ & high $r$ \\
\hline
related & 40 & 60 \\ \hline
unrelated & 95 & 5 \\
\end{tabular}
\end{center}
\end{table}%

From the table we see that the $p$-value of 5\% is a statement about the unrelated time-series.  It says nothing about the related time-series.  To get a statement about the related time-series we need to be able to repeat our experiment a sufficient number of times to produce a trustworthy probability density of the correlation values for related time-series.  This is often impossible.  Regularly we only have a single series, say from a climate record.  We can then not infer the probability density without extra information or some physically based estimates about the sizes and properties of the signal and the noise.

Based on this example table, we can now partly answer the question that most people are interested in: is the observed correlation $r_0$ an indication of a real relation or is it a fluke?  If we assume that the observed correlation is larger than the threshold correlation $r_p$ then we see from the above table that the chance of it being  representative of a real relation is $60/(60+5)\approx 92\%$, where we have employed equal prior odds on the time-series being related or unrelated; this probability is different from the 95\% that the significance test would have us believe.

Note that the 92\% value above depends on the prior odds.  If we do not know whether the time-series are related or unrelated, it does not mean these two options have equal odds; it just means that the odds are undefined, see Cox (1961).  The assumption of equal odds is a strong additional assumption, although it can be thought of as the maximum entropy prior, that is, it is the assumption that is maximally non-committal given lack of any further information regarding the relation between the time-series, see Jaynes (1963, 2003).  Of course, in reality such equal prior odds are unlikely, and it is usually impossible to quantify the actual prior odds.

The actual probability also depends on the division between the high and low probabilities for the related time-series.  If the signal--to--noise ratio is low in our experiments we expect a weak distinction between related and unrelated time-series.  In the limit of very low signal--to--noise ratio, the related series would also show 95\% low correlations and 5\% high correlations, see table below.
\begin{table}[h!]
\begin{center}
\begin{tabular}{r|c|c}
low signal/noise & low $r$ & high $r$ \\
\hline
related & 95 & 5 \\ \hline
unrelated & 95 & 5 \\
\end{tabular}
\end{center}
\end{table}%
The probability that our observed $r_0$ with $r_0>r_p$ is indicative of an actual relation is then $5/(5+5)=50\%$, again assuming equal prior odds for the time-series to be related or unrelated: the observed correlation does not provide evidence either way, even though it is thought to be `significant' according to a significance test.  Of course, this should not come as a surprise: if the signal--to--noise ratio is very low, then any observed correlation essentially provides information about the noise and it is therefore impossible to use this observation to infer anything about the signal.  Although this last case represents an extreme example, it does demonstrate that the $p$-value can be very far from the actual probability of the truth of a null-hypothesis.

\locsec{3. Bayesian analysis}

We can formalize the situation by using Bayesian statistics.  Let us define the hypothesis $H$ as `the time-series are related.'  We observe that the time-series have a correlation of $r_0$.  We are now interested in
the conditional probability, $p(H|r_0),$ that the hypothesis is true, given that the time-series have correlation of at least $r_0$.  The significance test gives us the $p$-value, that is, the conditional probability $p(r_0|\overline{H})$ that we observe a correlation $r_0$ given that the hypothesis is false ($\overline{H}$).  So:
\begin{equation}
p\text{-value} \equiv p(r_0|\overline{H}).
\end{equation}
It is important to keep this Bayesian expression for the $p$-value in mind.

A common mistake is to assume that $p(H|r_0)= 1-p(r_0|\overline{H}).$ This is the mistake of the transposed conditional: it is wrongly assumed that $p(r_0|\overline{H})=p(\overline{H}|r_0).$  It is straightforward to do the correct algebra:
\begin{align}
p(H|r_0) &= 1-p(\overline{H}|r_0) && \text{(complementarity)} \nonumber \\
         &= 1-p(r_0|\overline{H})\frac{p(\overline{H})}{p(r_0)} && \text{(Bayes' theorem)} \nonumber \\
         &= 1-p(r_0|\overline{H})\frac{p(\overline{H})}{p(r_0|H)p(H)+p(r_0|\overline{H})p(\overline{H})} && \text{(exclusive propositions)} \nonumber \\
         &= 1-p(r_0|\overline{H})\frac{1}{p(r_0|H)O(H)+p(r_0|\overline{H})}, && \label{post}
\end{align}
where we have introduced the (prior) odds ratio for the hypothesis $H$,
\begin{equation}
O(H) = p(H)/p(\overline{H}).
\end{equation}
This equation is essentially Bayes' theorem written out to indicate the relationship between the posterior probability $p(H|r_0)$ and the $p$-value.   With this equation it is obvious that we cannot use the $p$-value $p(r_0|\overline{H})$ to estimate the probability of the truth of the hypothesis.  We also need the prior odds ratio as well as the conditional probability $p(r_0|H)$.  Note that, if we assume an odds ratio of $O(H)=1$, then we recover the results we presented in the previous section.

Perhaps in hindsight, it should come as no surprise that the probability of the truth of $H$ needs to depend on the prior odds for $H.$  If $H$ is overwhelmingly likely ($O(H)\rightarrow\infty$), then the observation of correlation $r_0$ does very little to change this: $p(H|r_0) \rightarrow 1.$  If $H$ is very unlikely ($O(H)\rightarrow 0 $), then the observation of correlation $r_0$ does again very little to change this: $p(H|r_0) \rightarrow 0.$ 

It is also interesting to consider again the case of low signal--to--noise ratio.  In this limit the conditional probabilities $p(r_0|H)$ and $p(r_0|\overline{H})$ become indistinguishable.  From Eq.~\ref{post} we then find
\begin{equation}
p(H|r_0) \approx \frac{O(H)}{1+O(H)} = p(H).
\end{equation}
As expected, in this case the observation of $r_0$ changes nothing to the probability of $H$; the observed correlation is mainly a measure of the noise and says little about the signal.  For a prior odds ratio of 1, the probability for the hypothesis to be true remains 50\% after the observation.

Written out like this, it seems surprising that so many of us regularly get confused by significance tests at all.  Let us analyse the following apparently innocuous statements which in some form or another seem to be the mainstay of many investigations, for example a physical measurement:
\begin{enumerate*}
\item My measurement stands out from the noise.
\item So my measurement is not likely to be caused by noise.
\item It is therefore unlikely that what I am seeing is noise.
\item The measurement is therefore positive evidence that there is really something happening.
\item This provides evidence for my theory
\end{enumerate*}
The first two statements are essentially expressions of the fact that we have a situation with a low $p$-value: the chance that the observation is produced by noise is low.  The main error occurs in the third statement.  It is the error of the transposed conditional.  The probability of the data to be noise, given our measurements, is not the same as the probability of our measurements, given that the data is noise. The fourth statement would follow from the third statement if it were true.  The truth of the fifth statement depends on what alternatives there are to the noise hypothesis; this is where physics comes in as well as Occam's razor: is my theory the next most likely explanation of the observation?  The presence of alternative theories also influences prior odds for hypotheses.  For example, if there are many plausible alternative hypotheses, the present hypothesis will have low prior odds.  A beautiful quantification of such ideas can be found in Jaynes, 2003.

A more compact form of Eq.~\ref{post} can be found by writing Bayes' theorem in terms of prior odds ratio $O(H)$ and posterior odds ratio $O(H|r_0)$ with
\begin{equation}
O(H|r_0) = p(H|r_0)/p(\overline{H}|r_0).
\end{equation}
We find
\begin{equation}
O(H|r_0) = O(H)\,\frac{p(r_0|H)}{p(r_0|\overline{H})}.  \label{odds}
\end{equation}
The factor which updates the prior odds to the posterior odds is called the likelihood ratio.  For example, in the case of a low signal--to--noise ratio, the likelihood ratio equals 1; in this case the posterior and prior odds are the same.  Note again, that to find the posterior odds, the $p$-value, $p(r_0|H),$ is insufficient; we need the prior odds as well as the likelihood ratio.

So what do we do?
Equation~\ref{odds} gives some quantitative clues.  It is true that from Eq.~\ref{odds} it follows that a low $p$-value seems to indicate that the odds for $H$ typically have increased by our `statistically significant' observation.  By how much depends on the value of the likelihood ratio $p(r_0|H)/p(r_0|\overline{H}).$  If the $p$-value is low \textsl{compared to} $p(r_0|H)$ then the posterior odds for $H$ are larger than the prior odds.  Although its value is usually hard to determine, we normally assume that $p(r_0|H)$ is not small (it depends, for example, on the signal--to--noise ratio.)  In this sense a low $p$-value \textsl{can} provide positive evidence for the hypothesis.  What it does \textsl{not} provide is any quantitative measure of what the posterior odds are or by what amount the odds might have improved.  The 5\% (or 1\%) significance bound is utterly irrelevant: the improvement or deterioration of the odds for $H$ depend on how large the $p$-value is compared to $p(r_0|H),$ a quantity that in practice is hard to determine.

\locsec{4. Conclusions}

So are significance tests at all useful?  As indicated before, a high $p$-value is a useful indication that our observed correlation is not particularly noteworthy.  Note that a high $p$-value (that is, high $p(r_0|\overline{H})$) does not mean that $p(H|r_0)$ is low, see Eq.~\ref{post}.  It just means that the observed correlation is easily consistent with null-hypothesis $\overline{H}$ so that $\overline{H}$ cannot be rejected.  Occam's razor then tells us that we should not hypothesize a relationship for which there is no evidence.

Oppositely, a low $p$-value is not indicative of much at all except that the observed correlation is not very probable if the null-hypothesis were true.  There is a tentative, but unquantified and possibly incorrect, indication that the posterior odds for our hypothesis may have increased, as quantified by Eq.~\ref{odds}. But especially in this case, which is regularly used as positive evidence for the hypothesis, any information \textsl{assuming} the null-hypothesis is quite irrelevant.

A so-called `significant correlation' is meaningless in any practical sense; such a statement is a category error.  Significance tests of a single experiment alone cannot be used to provide quantitative evidence for a physical relation.

\textsl{Acknowledgments:~} The author thanks R. G. Harrison for insightful discussions during the preparation of this manuscript.

\locsec{References}

J. S. Armstrong, 2007: Significance tests harm progress in forecasting.\textsl{Int. J. of Forecasting}, \textbf{23}, 321-327

J. Cohen, 1994: The Earth is Round ($p<0.05$). \textsl{American Psychologist}, \textbf{49}, 997--1003.

R. T. Cox, 1961: \textsl{The Algebra of Probable Inference.} The John's Hopkins Press, Baltimore, 114pp.

J. E. Hunter, 1997: Needed: A Ban on the Significance Test, \textsl{Psychological Science}, \textbf{8}, 3--7.

E. T. Jaynes, 1968: Prior Probabilities, \textsl{IEEE Trans.\ on Systems Science and Cybernetics}, \textbf{4}, 227--241.

E. T. Jaynes, 2003: \textsl{Probability Theory.  The Logic of Science.} Cambridge University Press, Cambridge, 727pp

I. N. Jolliffe, 2004: P stands for \ldots, \textsl{Weather}, \textbf{59}, 77--79.

N. Nicholls, 2001: The Insignificance of Significance Testing. \textsl{Bull. Amer. Met. Soc.}, \textbf{82}, 981--986.

S. T. Ziliak \& D. N. McCloskey, 2008: \textsl{The Cult of Statistical Significance.} University of Michigan Press, 352pp. 




\end{document}